\documentstyle[12pt]{article}
\newcommand{\beq}{\begin{equation}}
\newcommand{\eeq}{\end{equation}}

\def\op{operator}

\def\cn{condition}

\def\rep{representation}

\begin{document}
\begin{titlepage}
\begin{flushright}
ITEP-67-1993\\
October 1993\\
\end{flushright}
\vspace{0.5cm}
\begin{center}
{\large {\bf Liouville Theory and Special Coajoint Virasoro Orbits}}\\
\vspace{1.5cm}
{\bf A.Gorsky}
\footnote{e-mail:gorsky@vxitep.itep.msk.su}\\
\vspace{0.4cm}
{\em Institute of Theoretical and Experimental Physics,\\
B.Cheremushkinskaya,
Moscow, 117259, Russia}\\
\vspace{0.4cm}
{\bf A.Johansen}
\footnote{e-mail:johansen@lnpi.spb.su}\\
\vspace{0.4cm}
{\em The St.Petersburg Nuclear Physics Institute\\
 Gatchina, St.Petersburg District, 188350 Russia}

\end{center}
\begin{abstract}
We describe the Hamiltonian reduction of the coajoint
Kac-Moody orbits to the Virasoro coajoint orbits explicitly in
terms of the Lagrangian approach for the Wess-Zumino-Novikov-Witten
theory.
While a relation of the coajoint Virasoro orbit
$Diff \; S^1 /SL(2,R)$ to the Liouville theory has been already
studied we analyse the role
of special coajoint Virasoro orbits
$Diff \; S^1/\tilde{T}_{\pm ,n}$ corresponding to
stabilizers generated by the vector fields with double zeros.
The orbits with stabilizers with single zeros do not appear
in the model.
We find an interpretation of zeros $x_i$ of the vector field of
stabilizer $\tilde{T}_{\pm ,n}$
and additional parameters $q_i$, $i = 1,...,n$,
in terms of quantum mecanics
for $n$ point particles on the circle.
We argue that the special orbits are generated by
insertions of "wrong sign"
Liouville exponential into the path integral.
The additional parmeters $q_i$ are naturally interpreted as
accessory parameters for the uniformization map.
Summing up the contributions of the special Virasoro orbits we get
the integrable sinh-Gordon type theory.

\end{abstract}
\end{titlepage}
\newpage
\section{Introduction}

The \rep s of the affine algebras and the Virasoro algebra
play the crucial role in the modern approaches to the 2D
conformal and integrable theories.
The physical \op s
are classified in terms of the \rep s and moreover the partition
functions can be found in terms of the characters of the \rep s
relevant
for a particular system.
A popular example is the partition function of the 2D Yang-Mills
theory
which can be expressed in terms of the characters
of the \rep s for the corresponding finite dimensional algebra which
classifies them for the affine case.
There is a rich structure for the \rep s of the Virasoro algebra.
Having in mind Kirillov-Konstant construction which
allows to find the \rep s of the algebra starting from the
quntization of
the coajoint orbits we can reduce the problem to the classification
of the orbits.
For the Virasoro algebra
the classification can be obtained in terms of
the stabilizer vector fields (see as a review ref.\cite{witten}).
There are four types of the stabilizers:
$S^1$, $SL_n(2,R)$, $\tilde{T}_{\pm ,n}$ and $T_{\Delta ,n} .$
First two types of the orbits admit a transparent interpretation:
they correspond to the \rep \ with the highest vector
for $S^1$ and the \rep s with the null vectors for $SL_n(2,R).$
The last two types of the orbits did not attract much attention of
physists.
The point is that when quantizing the corresponding orbits one deals
with
an unbounded Hamiltonian.
But these types of \rep s are interesting because they introduce
some non-trivial effects into the game.
Namely it was shown in refs.\cite{gor,khes}
that these orbits can be described
via the Hamiltonian reduction procedure starting from the affine
$SL(2,R)$ group if one performs the reduction of an element of
$SL(2,R)$ with a nontrivial winding number.

In what follows we shall discuss the Lagrangian approach
to the special Virasoro orbits.
A natural object associated with a coajoint orbit of the affine algebra
is the Wess-Zumino-Novikov-Witten (WZNW) action which
provides the symplectic structure on this orbit.
To describe the reduction procedure it is sufficient to couple the WZNW
theory to a gauge field.
In particular the reduction to the Virasoro algebra or to the associated
Liouville action results from the gauging of the Borel subgroup of the
affine algebra \cite{as,raif}.
We will try to argue that the special Virasoro orbits appear if one
takes
into consideration the classical background of the gauge field
which reminds the vortex configuration and has a non-trivial
winding number.
In this description each special coajoint orbit is natuarlly enlarged
to a finite set of orbits of singular 2-differentials
on $S^1$ with single poles which are stabilized by
regular vector fields
with double zeros.
We find an interpretation of residues of these poles in terms of
quantum
mechanics on $S^1$ and from the point of view of the uniformization
problem of Riemann surfaces.

The paper is organized as follows.
In sections 2 and 3 we will describe the reduction of the WZNW
theory to the geometric action which corresponds to the special
Virasoro orbits. In section 4 a possible relation to the
uniformization problem will be discussed.
In Conclusion
we collect main statments of the paper.

\section{Hamiltonian Reduction to the Special Virasoro Orbits}
\setcounter{equation}{0}

We consider the $SL(2,R)/B\times B$
coset conformal theory where $B\times B$ group corresponds to
the left and right rotations by the Borel subgroups of $SL(2,R)$.
This theory can be
described by the gauged Wess-Zumino-Novikov-Witten (WZNW)
theory with the following Lagrangian
\beq
{\it L} = L_{WZNW} -
\frac{k}{4\pi}  \bar{A} {\rm Tr}g^{-1} g' \;t_+  -
\frac{k}{4\pi} A {\rm Tr} t_- \; \dot{g} g^{-1} -
\eeq
$$-\frac{k}{4\pi} A \bar{A} {\rm Tr}t_- gt_+ g^{-1}
- \bar{\mu} A - \mu \bar{A}.$$
Here $k$ is a level of the Kac-Moody algebra,
$t_-$ and $t_+$ are the matrices
\beq
t_+ = \left( \begin{array}{cc} 0& 1\\ 0&0 \end{array} \right),\;\;
t_- = \left( \begin{array}{cc} 0& 0\\ 1&0 \end{array} \right).
\eeq
$A$ and $\bar{A}$ are the corresponding gauge fields,
$\mu$ and $\bar{\mu}$ are constant parameters and
$g'$ and $\dot{g}$ are the derivatives of the group element $g$ with
respect to the light-cone coordinates $x = x_1 + x_2$ and $t=x_1 -x_2,$
respectively
(in this section we consider the Minkovskian version of the theory
with respect to the 2D world sheet).
The light-cone
coordinate $x = x_1 + x_2$ is compactified
on a circle of unity radius, while the second coordinate
$t = x_1 - x_2$ is for definitness non-compactified.

It is convenient to use the Gauss parametrization
for the group element
\beq
g= (1 + \alpha t_-) e^{\sigma_3 \phi} (1 + \beta t_+).
\eeq
Here $\alpha, \; \beta$ and $\phi$ are real scalar fields.
This representation is not quite well defined on all the group
manifold.
Actually it is necessary to use four copies of such a
representation to cover all the group manifold  \cite{raif}.
However we can consider the part of the group manifold
where this representation is correct.

In terms of the fields $\alpha, \; \beta$ and $\phi$
the Lagrangian of the model reads as
\beq
{\it L} = - \dot{\phi} \phi' + e^{2\phi}(\alpha' + A)
(\dot{\beta} +\bar{A}) - \bar{\mu} A
- \mu \bar{A}.
\eeq
Integrating over the gauge fields $A$ and $\bar{A}$ we easily get
the Lagrangian of the Liouville theory
with the cosmological term (see, e.g. \cite{moore} and refs. there)
\beq
{\it L} = - \dot{\phi} \phi' - \mu \bar{\mu} e^{-2\phi}.
\eeq
We ignored here the volume of the gauge group $B\times B$.
The normalization of the measure in the functional integral is
determined by the metric on the group manifold
\beq
||\delta g||^2 = \int (- \delta \phi^2 + e^{2\phi} \delta \alpha \delta \beta).
\eeq
This fact should be taken into account when we calculate the quantum
corrections to the Lagrangian.

Now we want to interprete this Liouville theory in terms of the
coajoint Virasoro orbits.
To this aim we split the functional space of the gauge fields $A$
into classes of equivalence as follows.
We integrate over all $\bar{A}$ gauge fields with a fixed
$A$ field.
In this case the field $A$ can be considered as
a representative of a coajoint orbit of Kac-Moody algebra while
the gauging
of the Borel subgroup acting from the right can be considered as
the Hamiltonian reduction \cite{as}.
Actually in this situation we have
the Lagrangian
\beq
{\it L} =- \dot{\phi} \phi' - \bar{\mu} A,
\eeq
with the following constarint
\beq
\partial \alpha +A = \mu e^{-2\phi}.
\eeq
For convenience we omitted here the factor $k/4\pi$ which can be easily
inserted when it is necessary.

Now we need to make a comment on restrictions to the field space.
The group element $g$ is assumed to be
smooth and periodic with respect to the $x$ coordinate.
The point is that the classification of coajoint Virasoro orbits
corresponds to a splitting of the space of (2,0)-forms $b$ into
classes of equivalence with respect to smooth deformations
with a central extension, i.e. $Diff\; S^1$.
The \cn \ of the smoothness of the group element $g$ is
introduced because we want to include all the
homotopically non-trivial information into the field $A$ which can so
have singularities.
In this way we will classify the coajoint Virasoro orbits in terms of
homotopic equvalence classes of $A$.
The field $A$ is assumed to be periodic and single-valued but
it may have poles.

To make a reduction from the coajoint Kac-Moody orbit
corresponding to the representative $At_-$
to a Virasoro
coajoint orbit it is convenient to start once again from the beginning,
i.e. with eq. (2.1).
First we should put the matrix
$At_-$ into the following form
\beq
\tilde{A} = \left( \begin{array}{cc}
0& u\\ 1&0 \end{array} \right)
\eeq
by a Kac-Moody transformation
\beq
g \to h g,
\eeq
with a regular group element $h$.

Using the Polyakov-Wiegman formula it is easy to calculate the change of the
Lagrangian under the transformation with the group element $h$
\beq
{\it L} = L_{WZNW}
-\frac{k}{4\pi} {\rm Tr} \tilde{A} \dot{g}
g^{-1} -
\frac{k}{4\pi} {\rm Tr} \tilde{A}  g \bar{A} g^{-1} -
\frac{k}{4\pi} {\rm Tr} g^{-1} g' \bar{A}
- \frac{k}{4\pi}{\rm Tr} At_- \dot{h} h^{-1}.
\eeq
Here
\beq
\tilde{A}= h^{-1}  h' + h^{-1} At_- h.
\eeq
It is convenient to use
the Gauss representation for the group element $h$
\beq
h = (1+ a t_-) e^{\sigma_3 p} (1+ b t_+)=
\eeq
$$=\left( \begin{array}{cc}
e^p & b e^p\\
a e^p & ab e^p + e^{-p} \end{array} \right) .$$
Then it is easy to find the equations for the parameters $a$, $b$ and $p$
\beq
p' - b =0, \;
\;\; e^{2p} a' + e^{2p} A =1.
\eeq
The off-diagonal matrix element of $\tilde{A}$
corresponding to $u$ in eq.(2.9)
is a representative of the Virasoro coajoint orbit
\beq
u = p'' + (p')^2.
\eeq
Let us emphasize that the matrix elements of $h$ are assumed to be
smooth and periodic with respect to $x$.
However it is easy to see that it is not possible in general to put
the Kac-Moody representative $At_-$ into the form (2.9) for
a generic 1-form $A$.
Actually a short analysis shows
that it can be done only if the function $A$
does not have any poles of order higher than 1.
Therefore one can interprete this theory in terms of
the coajoint Virasoro orbits only if
we limit the functional space of $A$'s to contain fields
not very singular, i.e. with poles at most of order 1.
It is interesting that namely this constraint
corresponds to (almost)
finite action (2.7) for non-vanishing cosmological constant.
Indeed with an appropriate regularization of a singularity
in $\alpha$ the
cosmological term in eq.(2.7) has only logarithmical divergence:
\beq
\int A =\int (e^{-2\phi}- \alpha' )  .
\eeq
We also see that the contributions of
fields $A$ with stronger singularities are suppressed due to
a stronger divergence of classical action.

It is also clear that the gauge field $A$ does not have any zeros since
the exponential $\exp \; p$ is regular.

However a problem appears now that the representative of Virasoro
orbit in eq. (2.15) can be singular.
Indeed let $A = h/x$ where $q$ is a non-zero constant near the pole
$x =0$.
Then let us find an appropriate solution to eq.(2.14)
near $x=0$
\beq
p = \frac{1}{2} ln\; \nu + ln \;x - qx/2 =...,\;\;
a = \frac{r}{x} (1 + s x) +...,\;\;
b= \partial p = \frac{1}{x} - q/2 +...,
\eeq
where $\nu$, $h$, $r$, and $s$ are real parameters, $\nu > 0 .$
{}From eqs. (2.14) it is easy to see that
\beq
q = h\nu .
\eeq
In turn a direct calculation shows that
\beq
u = \frac{q}{x}.
\eeq
Moreover $ {\rm exp} (-2p)$ has no zeros since we assume that
$\exp \; p$ is regular.
Hence the poles of the 2-differential $u$ are determined only
by poles of $A$.

Now we want to show that in a sense
such a representation
corresponds to a special coajoint Virasoro
orbit with a stabilizer generated by a vector field with
double zeros.
Indeed, the coajoint
action of the Virasoro group (see, e.g. ref. \cite{witten})
reads
\beq
Ad*(F) (u(x),c) = (u(F(x))\cdot F'^2 (x) - (c/24\pi ) S(F) ,c),
\eeq
where $c$ is a central charge, $F$ is an element of $Diff\; S^1$
and $S(F)$ is the Schwartzian derivative
\beq
S(F) = \frac{F'''}{F'} -\frac{3}{2} \left( \frac{F''}{F'} \right)^2.
\eeq
Let us consider the equation for a stabilizer near
$x =0$ where the gauge field $A$ has a pole $A \propto 1/x$.
The Virasoro representative is $u = q/x$.
We have
\beq
-\frac{1}{2} \epsilon''' - \frac{2q}{x} \epsilon' +
\frac{q}{x^2} \epsilon =0.
\eeq
A regular solution to this equation (up to a normalization) is
\beq
\epsilon = x^2 - q x^3.
\eeq
Therefore $\epsilon $ can not have a single zero at this point.
Instead we get {\it a double} zero of the
stabilizer in the same point where
the gauge field $A$ has a pole.
It is worth to noticing that at $q \neq 0$
$\epsilon'''$ is not zero at zeros of $\epsilon$
in contrast to usual description of
the special $Diff\; S^1/\tilde{T}_{\pm ,n}$
Virasoro orbits \cite{witten}.

It is necessary to emphasize here that we did not claim
that the vector field $\epsilon$ of stabilizer has no
additional zeros.
Here we only point out that there are no
single zeros of the vector field $\epsilon$.
Indeed the stabilizing vector field is given by
$\epsilon = \exp 2p$ and hence a single zero would lead to
a double pole in the function $u$ without any single pole
at the same point.
In turn according to eq.(2.15)
this would lead to an appearence of a single pole
in the field ${\rm exp} (-2p)$ which can not of course have a
single pole without a double pole at the same point
(this is a consequence of the regularity of the group element
$h$ in eq.(2.12)).
Therefore we conclude that the vector field $\epsilon$ can not have a single
zero, and hence the orbits $Diff \; S^1/T_{\Delta ,n}$
do not appear in this model.
As to $Diff \; S^1/SL_n (2,R)$ we shall see in next section that
these orbits can appear while $Diff \; S^1/S^1$ do not enter the model.

Another important observation is that if we consider a limit
of vanishing residues we have still the vector field $\epsilon$
with double zeros that is a stabilizer of a regular 2-differential
representing the special Virasoro orbit with a stabilizer
$\tilde{T}_{\pm ,n} .$
Therefore we may consider these special Virasoro orbits as a limit
of vanishing residues.

Notice that the residue of the pole in the Virasoro representative $u$
is not uniquely defined even if the residue in the gauge field $A$ is fixed.
The freedom for a choice of the residue in the
2-differential $u$ corresponds
to a possibility to change this residue by diffeomorphisms while
it is not possible for 1-form $A$.
However no diffeomorphism with positive orientation can change the sign
of the residue.
Hence the space of all orbits of 2-differentials
corresponding to stabilizers $\epsilon$
with $n$ double zeros can be splitted into $2 \cdot
3^n$ subspaces classified
by sets of $n$ numbers ($0,\; +1$ or $-1$)
(an additional doubling corresponds to $\pm$ signature of
a special orbit $Diff \; S^1 /\tilde{T}_{n,\pm}$).
These classes are not
intersecting and are reduced to the special Virasoro orbits (with
regular representatives) in the limit of vanishing residues.
Moreover such a singular 2-differential is
additionally characterized
by $n$ real numbers because the value
$\epsilon'''/(\epsilon'')^2$ at a zero of the vector field $\epsilon$
is invariant under the action of $Diff\; S^1 .$

\section{Geometric Action for the Special Virasoro Orbits}
\setcounter{equation}{0}

Now we want to put the action into the form which is most close
to the form of the geometrical action for the Virasoro orbit.
To this aim we change a parametrization of the functional space.

Let us come back to the Lagrangian (2.7).
Since the gauge field can have only single poles
the residues of the double poles in
the exponential $\exp \phi$ and in the 1-form $\alpha'$
should be correlated so that to satisfy eq.(2.8).
Therefore we can identify the pair of fields $(\alpha , \phi)$ with
the pair $(\alpha , p)$ (eq. (2.14)).

We can introduce the following parametrization of the space of fields $A$.
Let us choose a particular representative $A_0$ of an orbit
of gauge fields $A$.
Then for this field $A_0$ we can choose a pair of fields $\alpha_0$
and $\phi_0$ which provide us with a 2-form $u(x)$
representing the corresponding coajoint Virasoro orbit.
The existence of this pair is guaranteed by the constraint (2.8).

Taking into account that ${\rm exp}(-2\phi)$ is 1-form
we now can get any point of the Virasoro orbit using
the following equation
\beq
e^{-2\phi(x)}= F'(x) \; e^{-2\phi_0(F(x))},
\eeq
where $F$ is a diffeomorphism of $S^1 .$
This new field $\phi$ corresponds of course to the gauge field
$A(x)$ given by
\beq
A(x) = F'(x) A_0(F(x)).
\eeq
Therefore integrating over smooth fields
$F$ we cover all space of fields related to $A_0$, $\alpha_0$ and $\phi_0$
by the action of $Diff\; S^1 .$

Actually the equation (2.8) has generically an
infinite set of solutions for a given field $A.$
First, there is a continuous
degeneracy which corresponds to the gauge invariance of the theory
under transformation
\beq
A\to A + f' \;\;\; \alpha \to \alpha - f
\eeq
with a regular function $f$.
However this gives simply the volume of the gauge group
in the path integral.
Second, there is a discrete degeneracy which is discussed below.

Let us first consider the sector of the gauge fields $A$ without
poles.
In this case the field $A$ can be represented in the following form
\beq
A= <A> F' ,
\eeq
where $F$ is a diffeomorphism which can be represented as
\beq
F = f + x ,
\eeq
where $f$ is a regular periodic function.
If $\alpha$ is a regular periodic function
then after a rescaling of the field $\alpha$ and a shift of
$\phi$ by a constant we get from eq. (2.8)
\beq
\tilde{F}' = e^{-2\phi},
\eeq
where a diffeomorphism $\tilde{F} =\alpha +f + x$
obeys the same boundary condition as $F,$
$\tilde{F} (x +2\pi) =\tilde{F} (x) +2\pi .$
It is clear that
this case corresponds to the coajoint Virasoro orbit
$Diff\; S^1/SL(2,R)$ with a zero representative \cite{as}.
Actually there is an infinite set of solutions to eq.(2.8) for
the gauge field (3.4).
One can show that they are parametrized by the number of zeros of the
exponential $\exp \phi .$
It is easy to find, for example, the following solution to eq.(2.8)
\beq
\phi_0 = \ln \;\cos\; \frac{nx}{2}
 -\frac{1}{2} {\rm ln}
\;(1+ \cos^2 \frac{nx}{2}),
\eeq
while
\beq
\alpha_0 = \frac{2}{n} \tan \; \frac{nx}{2} .
\eeq
It can be verified that this orbit of the field $\phi$ corresponds
exactly to the Virasoro orbit with
a stabilizer with $n$ double zeros.
This choice of the field $\phi_0$ corresponds to
\beq
u = - \frac{n^2}{2} \frac{1 +\sin^2 \frac{nx}{2}}{(1 +
\cos^2 \frac{nx}{2})^2}.
\eeq
In turn we can try to find a regular solution to eq. (2.15) for
$u = constant$.
Then we easily see that such a solution exists only for $u = -n^2/4$ where
$n$ is an integer and has the following form
\beq
\phi_0 = \ln \; \cos \; \frac{nx}{2} .
\eeq
This solution of course corresponds to the vanishing gauge field $A$
and hence gives no contribution to the dependence of
the
Liouville partition function (for the Lagrangian (2.7))
on the
cosmological constant $\mu\bar{\mu} .$
It is also clear that
this Virasoro orbit has the stabilizer $SL_n (2, R)$ \cite{witten}
which is generated by a vector field $\epsilon$ without zeros
and two vector fields with $n$ single zeros.
It is worth emphasizing that this the only type of orbits with
a constant representative.

Notice that the singular fields $\phi_0$ of the type of those in
eqs.(3.7) and (3.10) are in fact reduced $SL(2,R)$ vortices as
it was discussed in ref.\cite{gor}.
To check this one should substitute the group element
$g = \exp inx\sigma_2$
into eq.(2.3) and then rotate it into the form (2.9).

Thus it is clear that the space of solutions to eq.(2.8) for the
regular fields $A$ covers
an infinite set of orbits with stabilizers with double zeros.
Actually as it is shown in the previous section each special Virasoro
orbit associated with a stabilizer with $n$ double zeros
corresponds to $n$-dimensional space of (singular) 2-differentials.
The singular 2-differentials
are shown to correspond to gauge fields $A$
with single poles.
It is clear that for a given gauge field $A$ there is an infinite set
of gauge non-equivalent solutions of eq.(2.8).
These are parametrized by the number of double zeros
$x_i$, $i=1,...,n$ of a stabilizer
$\epsilon$ and $n$ parameters $q_i$ defined as the residues of
2-differential associated with the vector field $\epsilon .$

The considerations above for static fields can also be generalized
to the fields depending on both light cone coordinates $x$ and $t.$
In this situation we should consider diffeomorphisms
$F(t ,x)$ depending on "time", i.e. on $t.$
A representative $\phi_0 (t ,x)$ of the orbit of the field $\phi$
can depend in general on $t$ explicitly.
Then the variables of integration in the path integral are not exausted
by diffeomorphisms $F(t ,x)$ and we should assume that
the positions of zeros $x_i$ of a stabilizer and the parameters $q_i$
are also variables of integration.
Thus we can introduce a dynamics on the space of singular
2-differentials.

To express the action in terms of new parametrization of functional space
we can make a shift of the field $\phi$ in the equation above
and hence in the Lagrangian (2.7) by $2\phi_0(t,F(t ,x))$.
Then we get an expression for the Lagrangian in terms of
a diffeomorphism $F$
\beq
L = -\phi' \dot{\phi} - u(t ,F)\; F' \dot{F}
- \dot{\phi_0}(t,F) \phi'_0 (t,F)\; F' -\bar{\mu} A(t ,x),
\eeq
where the smooth
fields $\phi$ and $F$ are connected by the following constraint
\beq
F' = e^{-2\phi}.
\eeq
This Lagrangian has a form which is very close to the form of geometrical
action \cite{as}.
In particular for $u=-n^2/4$ ($n \in {\bf Z}$) and $\phi_0$ given by
eq.(3.10) this lagrangian gives the geometrical action
for the $Diff\; S^1/SL_n (2,R)$ Virasoro
orbit \cite{as}
\beq
L = - \phi' \dot{\phi} + n^2 F'\dot{F}/4 - \mu\bar{\mu} <A>/2\pi.
\eeq
The third term in eq.(3.11) vanishes for this case.

It is necessary to emphasize however
that the Lagrangian above contains also two additional
terms as compared to the ordinary geometrical action \cite{as}.
The remarkable fact is that
both of them do not actually depend on the quantum field $F$
when integrated over $S^1$ and hence characterize only
a choice of the orbit of the gauge field $A_0$.

It is easy to see that one may consider the values $x_i$
of positions
of zeros of the stabilizer and the residues $q_i$ of poles
in $u(t,x)$ as quantum mechanical variables,
while the space of fields $F$ can be restricted to the
diffeomorphisms which
do not change the parameters $x_i$ and $q_i .$
Thus we get a quantum mechanics of n particles on a circle $S^1$
coupled
to the quantum field theory.

To understand the meaning of the variables $x_i$ and $q_i$
one can try to extract an effective quantum mechanical Lagrangian
for $x_i$ and $q_i$ integrating over field F in semiclassical
approximation at $k\to \infty$
(the Kac-Moody level $k$ is the coupling constant).
However there is no regular solution to the classical equation of
motion
for the field $F .$
Instead a configuration can be found that asymptotically
satisfies the equation of motion for the field $F .$
Such a configuration is equivalent to $F(x,t) = x$
with a particular choice of the classical background field $\phi_0$
which
is concentrated near the positions of singularities.
If $\phi_0$ is suppressed (for example, exponentially)
at $|x-x_i|< \delta$ (where $\delta << 1$ is a parameter)
then to the leading approximation the quantum mechanical action reads
as
\beq
L_{QM} \sim \delta \sum_i \dot{x}_iq^2_i .
\eeq
Thus one can see that the variables $x_i$ and $q_i$ are dual each
other.

\section{Special Virasoro Orbits in the Liouville Theory}
\setcounter{equation}{0}

Now we want to find a relation of the considerations
above to the Liouville theory formulated on the euclidean worldsheet.
With this modification of the formulation of the theory the
positions of singularities of the fields
in the chiral sector are rather complex points
on the disk.
That means that in the antichiral sector there are singualrities in the
complex conjugated points.

Correspondingly the formulas related to the special Virasoro
orbits are modified when translated to the euclidean formulation.
In particular,
near a singularity at $z=z_i$ of the classical field $\phi_0$ we have
\beq
\phi_0 =\frac{1}{2} {\rm ln} \nu_i + {\rm ln} |z-z_i|^2
+{\rm Re} h_i (z-z_i) ,
\eeq
where $z$ is a complex
coordinate on the disk,
$\nu$ is a real number, $\nu >0$, and $h_i$ is
a complex constant.
In turn the classical configuration $\phi_0$ determines a
meromorphic 2-differential
$u=(\partial \phi_0)^2 + \partial^2 \phi_0$
($\partial$ stands for the derivative in $z$).
This 2-differential is stabilized by a holomorphic vector
field $\epsilon = \exp 2\phi_0 .$
Instead of $Diff\; S^1$
one has to consider here the algebra of conformal transformations.
However if we draw a smooth non-self-intersecting
contour $K$ through the singularities of $\phi_0$ then
the generators of conformal transformations restricted to this contour
correspond to certain generators of the
complexified algebra of vector fields on $S^1 ,$
$c\; Vect \; S^1 ,$
while the contour $K$ can be understood as an image of $S^1$ under
a holomorphic function on a unit disk \cite{kirillov}.
This translation of the vector field $\epsilon =
\exp 2\phi$ gives an element of $c\; Vect \; S^1$
with double zeros.
Unfortunately such a correspondence does not give yet
any exact description
of the special Virasoro orbits in terms of the Liouville theory
on a complex disk.
Nevertheless we may assume that such a relation exists.

Let us now consider the Liouville theory.
As it is known
the quantum corrections coming form the integration measure
in the path integral (eq.(2.6)) modify the lagrangian
\cite{verlinde}.
Therefore at the quantum level we have
\beq
L = \sqrt{\hat{g}} \left( \frac{1}{2\pi \gamma^2}
(\hat{\nabla} \phi)^2 - \frac{Q\gamma}{4\pi} \phi R(\hat{g}) \right)
+ \frac{\mu\bar{\mu}}{8\pi \gamma^2}
\sqrt{\hat{g}} e^{-2 \phi}.
\eeq
Here we introduced a background metric on the Riemann surface.
The parameter $Q$ is related to the central charge of the theory
by equation
\beq
c=1+3Q^2, \;\;\;Q= \frac{2}{\gamma} + \gamma.
\eeq
In 2D gravity \cite{moore}
the parameter $\gamma$ is determined by a requirement that
the cosmological constant is represented by a marginal \op \ at the
quantum level. In our case its value follows from
the \cn \ that the theory is conformally invariant \cite{verlinde},
so that
\beq
Q= (k-1) \gamma ,\;\;\; \gamma = \sqrt{2/(k-2)} .
\eeq
Notice that this value of a parameter $\gamma$ corresponds to the
correct branch of solution to eq.(4.3)
(this is defined by the condition that
a semiclassical limit of the theory corresponds to
$\gamma \to 0$ \cite{moore})
\beq
\gamma = \frac{Q}{2} - \frac{1}{2} \sqrt{Q^2 -8}
\eeq
only if $k >3$,
while for $2<k<3$ the parameter $\gamma =\sqrt{2/(k-2)}$ obeys
\beq
\gamma = \frac{Q}{2} + \frac{1}{2} \sqrt{Q^2 -8}.
\eeq
After a rescaling $\phi \to -\gamma \phi /2$ we come to the
standard expression for the Lagrangian
\beq
L = \sqrt{\hat{g}} \left( \frac{1}{8\pi}
(\hat{\nabla} \phi)^2 + \frac{Q}{8\pi} \phi R(\hat{g}) \right)
+ \frac{\mu \bar{\mu}}{8\pi \gamma^2}
\sqrt{\hat{g}} e^{\gamma \phi}.
\eeq
The $zz$ component of the
energy momentum tensor for this theory reads as ($T_{z\bar{z}} =0$ since
the theory is conformal invariant)
\beq
T_{z z} =-\frac{1}{2} (\partial \phi)^2 + \frac{1}{2} Q
\partial^2 \phi .
\eeq
It is easy to that the stress tensor given above
is actually a modification (by quantum corrections)
of the expression for the representative of
the Virasoro orbit (2.15).

Now we want to use this formula to
interprete the special Virasoro orbits as a presence of insertions
of certain operators into the Riemann surface.
In turn such an interpretation is related
with the problem of uniformization (see, e.g. ref.\cite{moore}).

As it is known the uniformization map ${\cal J}$ of the Riemann sphere
$\bar{{\bf C}} = {\bf C} \cup \{ \infty \}$ with $n$ insertions of
operators $\exp \alpha_i \phi$ at points $z_i$, $i=1,...,n$,
can be determined by a ratio of
solutions of the Fuchsian differential equation
\beq
\frac{d^2 y}{dz^2} + \omega_X \; y =0,
\;\;\; \omega_X = \frac{1}{2} S[{\cal J}^{-1} (z); z],
\eeq
where $S[{\cal J}^{-1} (z); z]$ is the Schwarzian derivative.
The inverse uniformization map ${\cal J}^{-1}$ maps the
hyperbolic plane $H$ onto the Riemann sphere $\bar{{\bf C}}$ so that
$\bar{{\bf C}} \simeq H/\Gamma$, where $\Gamma$ is the Fuchsian group
uniformizing $\bar{C} .$

Near the positions of the inserted operators we have
\beq
\omega_X = \frac{\gamma^2}{2} T(z) =
\sum_i \left( \frac{\Delta_i}{(z -z_i)^2} +
\frac{c_i}{z-z_i} + {\rm regular\;
terms} \right)
\eeq
where
\beq
\Delta_i = \frac{1}{4} (1 - (1-\theta_i)^2),
\;\;\;\theta_i = \alpha_i \; \gamma ,
\eeq
and $c_i$ are the accessory parameters.

The exponential $\exp \gamma\phi$ is related to
the uniformization map ${\cal J}^{-1}$ \cite{tak}
\beq
e^{\gamma\phi(z)} = \frac{|({\cal J}^{-1})' (z)|^2}{({\rm Im}
{\cal J}^{-1} (z))^2},
\eeq
so that
\beq
T (z) = \frac{1}{\gamma^2} S[{\cal J}^{-1} (z); z] .
\eeq
In the semiclassical limit $\gamma \to 0$ taking
$\phi = -\gamma\phi_0 /2$ (where $\phi_0$ is defined by eq.(4.1))
we get from eq.(4.8)
\beq
T_{zz} = \frac{4h_i/\gamma^2}{z-z_i}.
\eeq
Comparing this expression with eq.(2.19) we formally see that
that our expression for the classical stress tensor for
the case of special Virasoro orbits can be interpreted as an effect
of inserted operators with $\alpha_i = 2/\gamma$
\beq
\prod_i e^{(2/\gamma) \phi(z_i)},
\eeq
while $2h_i$ play the role of accessory parameters.
In other terms we see that in the semiclassical limit the singularities
in the classical field $\phi_0$ are generated by the $\delta$-functional
sources $\sum_i \delta^2 (z-z_i)$ for the field $\phi .$
Actually it is easy to check that these operators are also responsible
for generating of the classical configuration $\phi_0$
(eq.(4.1)) at the quantum level (in the model without
the cosmological operator).
However for finite values of $\gamma$ the energy momentum
tensor (4.6) with $Q$ defined by eq.(4.3) has double poles at the
singularities $T_{zz}= 1/(z-z_i)^2 +....$

Notice that the parameter $\alpha$ obeying $\alpha \gamma= 2$
is actually another solution to the equation relating $Q$ and $\gamma$
\beq
\alpha = \frac{Q}{2} + \frac{1}{2} \sqrt{Q^2 -8} ,
\;\;\; {\rm at}\;\; k>3 ,
\eeq
and
\beq
\alpha = \frac{Q}{2} - \frac{1}{2} \sqrt{Q^2 -8} ,
\;\;\; {\rm at} \;\; 2<k<3 .
\eeq
Hence the \op \ $\exp (2\phi /\gamma)$ is also marginal.

Actually such an identification is not quite correct because the field
$\phi_0$ does not satisfy the classical
equation of motion with
sources in the presence of the cosmological term $\exp \gamma
\phi .$
Moreover the expression for $\phi_0$ does not satisfy to the boundary
conditions at singularities fixed in refs. \cite{tak}.
In turn in the case of insertions of operators
$e^{(2/\gamma) \phi(z_i)}$ there is no solution to the
classical equation of motion in the presence of the cosmological term
\cite{moore}.
Therefore such a classical configuration is stable only
if there is no
cosmological term in the lagrangian.

On the other hand at $k>3$ this situation
corresponds to the case when all the parameters are
$\theta_i = \alpha_i \gamma= 2$,
i.e. higher than the Seiberg bound \cite{seiberg}
$\alpha = Q/2$.
Recall that
this limiting value of parameter $\alpha$ is determined by the \cn \ that
the physical state corresponding to this \op \ can exist in the theory.
In particular such an \op \ does not allow any smooth semiclassical limit
$D \to -\infty$ in the two-dimensional gravity, where $D$ is
a central charge of `matter'.
Therefore the \op s ${\rm exp}(2\phi /\gamma)$
responsible for an appearence of
the special orbits are rather `non-physical'.

In the presence of the cosmological \op \ $\exp \gamma\phi$
in eq.(4.5) the semiclassical approximation for the correlators of
the \op s $\exp 2\phi/\gamma$ is not well defined.
However we conjecture here that at the quantum level
the special Virasoro orbits
correspond to insertions of the `wrong' sign Liouville exponential
(eq.(4.16)) at $k>3$,
while for $2<k<3$ the special orbits generate the usual
cosmological operator (eq.(4.17)).
Assuming the validity of the conformal Ward identities \cite{lat}
with insertions of the `wrong' sign exponentials we get
the quantum analog
of 2-differential of the special Virasoro orbit given by an expression
without double poles at positions of the operators
\beq
<T(z)\prod_i e^{(2/\gamma) \phi(z_i)}> =
\eeq
$$=\sum_{i=1}^{n-3}
\left(\frac{1}{(z-z_i)^2} +
\frac{1}{z-z_i} + \frac{z_i -1}{z} -\frac{z_i}{z-1} \right)
\frac{\partial}{\partial z_i} \ln <\prod_i e^{(2/\gamma)
\phi(z_i)}>,$$
where $SL(2,{\bf C})$ symmetry is assumed to be fixed by fixing
points $z_n$, $z_{n-1}$ and
$z_{n-2}$ to be equal to $0, \;1$ and $\infty .$
This also implies that the values of accessory parameters are
fixed by the positions of inserted operators and presumably
corresponds to `vacuum' values of residues $q_i$ in terms of
geometric quantization in previous sections.
{}From this point of view the parameters $h_i$ in eq.(4.1)
are rather the variables of integration in the path integral
near the classical configuration.

It is worth noticing that
the properties of the `unphysical' \op \ $\exp (2\phi/\gamma)$
are softer here
as compared to the usual Liouville theory.
The point is that in the usual Liouville theory
an insertion of \op \ with $\alpha > Q/2$
induces a curvature source for which the cosmological term is not
integrable, since it gives $\int 1/|z|^{\theta}$.
In the present formulation of the theory it is strictly
speaking not the case because the cosmological \op \
is proportional to $\int A = \int \partial \alpha + {\rm exp} (-2\phi)$
(at least at the classical level). The restrictions to singularities
of the fields $\phi$ and $\alpha$ are correlated so that
the integral can have only logarithmical divergence as in the kinetic term.

It is interesting to sum up contributions due to arbitrary insertions
of the `wrong' sign exponentials to the path integral for the
Liouville theory.
In the absence of the cosmological operator
this would correspond to a Coloumb gas.
Such a model is very similar to
the case of sine-Gordon model generated by contributions of vortices
in the theory of real scalar field compactified to $S^1 .$
However in contrast to the latter model here all `vortices'
are of the same charge since they are generated by the same operator.
Thus summing up the
contributions of special Virasoro orbits
the `wrong' sign Liouville exponential appears
in the lagrangian. Taking into account a cosmological term we
get a gravitational version of the
integrable sinh-Gordon model with a background charge
\beq
L = \sqrt{\hat{g}} \left( \frac{1}{8\pi}
(\hat{\nabla} \phi)^2 + \frac{Q}{8\pi} \phi R(\hat{g}) \right)
+ \frac{\mu \bar{\mu}}{8\pi \gamma^2}
\sqrt{\hat{g}} e^{\gamma \phi}+
\frac{\tilde{\mu}}{8\pi \gamma^2}
\sqrt{\hat{g}} e^{2\phi/\gamma},
\eeq
where $\tilde{\mu}$ is a parameter fixed by initial \cn s.
This Lagrangian is of course obtained in an approximation when
the interaction between both exponentials is not taken into account.
In the case $2<k<3$ and $\mu\bar{\mu} =0$ one can see that
the special orbits generate the usual cosmological operator.

At $k=3$ the solutions (4.16) and (4.17) coincide: in this case
the Virasoro central charge of the Liouville theory $c=25$ and
hence it correponds to $c=1$ two-dimensional gravity.
It is tempting to assume that the value $k=3$ would correspond
to a phase transition in the gas of the wrong branch (at $k>3$)
vertex operators
$\exp (2\phi/\gamma)$ since at this point the branch in
eq.(4.4) changes.

\section{Conclusion}

It was shown that gauging Borel subgroup in the WZWN theory  and
taking into account singular configurations of the gauge field  we
naturally get the structures assoshiated with the special Virasoro
coadjoint orbits.
Quadratic differentials with the single poles
appear when the residues in the poles of the gauge field do not
vanish.
Thus while the orbits with the stabilizers $\tilde{T}_{\pm ,n}$ can
be thought of as the perturbations of orbits with
$SL_n(2,R)$ ones the
orbits with the singular representatives results from the
additional perturbation by the singular gauge configurations.

It is known \cite{bow} that the
manifold $Diff\; S^1/S^1$ is  relevant  in
the closed string theory because of its close connection with  the
universal Teichmuller space of the Riemann surfaces with genus
$g>1 .$
The consideration above leads to an assumption that the parameters
of the singular 2-differentials can be interpreted as the
coordinates on the moduli space of the spheres with the marked
points. In this case the orbits with
$\tilde{T}_{\pm ,n}$ stabilizers
corresponding to vanishing residues are related with the
submanifold of moduli space with with zero "momenta".

We can not add something to the open problem of quantization
of the special coadjoint orbits. Standard methods are not
applicable here so new approaches should be developed.
But having
in mind the relations with the moduli space for the surfaces with
the marked points we can conjecture that the
conformal blocks for the
correlators on the sphere in the Liouville theory and
the related tau
functions are relevant objects for the description of the Hilbert
space in this hypothetical quantization.

Finally we want to mention a possible connection of our results
with dynamics of Witten's black hole which can be described as
a gauged $SL(2,R)/SO(1,1)$ WZNW model \cite{black}.
In this model there are vortices
corresponding to non-trivial winding numbers
(see, e.g. ref. \cite{verlinde}) and we can study
a problem of renormalization group flow induced by a gas of
such vortices.
On the other hand the theory considered in the present paper
correspond to a factorization of $SL(2,R)$ by a Borel subgroup
and as we see the singularities corresponding to the special
Virasoro orbits are in fact generated by vortices in
the group element of $SL(2,R) .$
Therefore we can conjecture that vortices in the
$SL(2,R)/SO(1,1)$ model would generate the `non-physical'
operators corresponding to the special Virasoro orbits.

We thank N.Nekrasov, P.Nelson, A.Rosly and K.Selivanov for the
useful discussions. A.G. thanks UBC and Aspen center for physics
where the part of this work was done for the kind hospitality.
A.J. is grateful to Fermi National Accelerator Laboratory
and Rutherford Appleton Laboratory
where the part of this work has been done for the warm hospitality.
This work was also supported, in part, by a Soros Foundation
Grant awarded by the American Physical Society.

\end{document}